\documentclass[prb,twocolumn]{revtex4}

\usepackage{amsmath,amssymb}

\usepackage{graphicx}

\begin{document}

\widetext

\title{Localization of Large Polarons in the Disordered Holstein Model}

\author{Oliver Robert Tozer$^{1,2}$\footnote{E.mail address: oliver.tozer@chem.ox.ac.uk} and William Barford$^{1}$\footnote{E.mail address: william.barford@chem.ox.ac.uk}}

\affiliation{$^1$Department of Chemistry, Physical and Theoretical
Chemistry Laboratory, University of Oxford, Oxford, OX1 3QZ, United
Kingdom\\
$^2$University College, University of Oxford, Oxford, OX1 4BH, United Kingdom}

\maketitle



\begin{center}
\large{{Abstract}}
\end{center}
We  solve the disordered Holstein model via the DMRG method to investigate the combined roles of electron-phonon coupling and disorder on the localization of a single charge or exciton. The parameter regimes chosen, namely the adiabatic regime, $\hbar\omega/4t_0 = \omega'  < 1$, and the `large' polaron regime, $\lambda < 1$, are applicable to most conjugated polymers. We show that as a consequence of the polaron effective mass diverging in the adiabatic limit (defined as  $\omega' \to 0$ subject to fixed $\lambda$) self-localized, symmetry breaking solutions are predicted by the quantum Holstein model for infinitesimal disorder -- in complete agreement with the predictions of the Born-Oppenheimer Holstein model.
For other parts of the ($\omega'$, $\lambda$) parameter space, however, self-localized Born-Oppenheimer solutions are not expected. If $\omega'$ is not small enough and $\lambda$ is not large enough, then  the polaron is predominately localized by Anderson disorder, albeit more than for a free particle, because of the enhanced effective mass.  Alternatively, for very small electron-nuclear coupling ($\lambda \ll 1$) the disorder-induced localization length is always smaller than the classical polaron size, $2/\lambda$, so that disorder always dominates. We comment on the implication of our results on the electronic properties of conjugated polymers.




\vfill
\pagebreak

\section{Introduction}

Electron-phonon coupling is believed to play a significant role in determining the electronic and optical properties of conjugated polymers. On a singly doped polymer this coupling causes local nuclear displacements around the doped charge. The charge and associated nuclear displacements, first proposed by Landau\cite{Landau}, is called a polaron, and the formation of a polaron is called `self-trapping'\cite{footnote1}. For translationally invariant systems the polaron forms a translationally invariant Bloch state. Similarly, Frenkel excitons -- created by the photoexcitation of a neutral polymer -- also couple to the nuclei and form exciton-polarons.

Conjugated polymers, however, are rarely translationally invariant: conformational and environmental disorder introduces another important effect in determining their electronic properties, namely the Anderson localization\cite{Anderson} of charges and excitons. Particles are always localized by disorder in one-dimensional systems\cite{Mott}. However, electronic states in the low-energy Lifshitz tail of the density of states are super-localized: these states are essentially nodeless and nonoverlapping, and have been termed `local ground states'\cite{Malyshev,Makhov}. Local charge ground states determine the donor and acceptor segments in conjugated polymers, while local exciton ground states determine the  spatial extent of chromophores (namely, the irreducible segments of a polymer chain that absorb or emit light)\cite{Barford2013}.

In this paper we describe the combined effects of electron-phonon coupling and disorder. We do this via an investigation of the one-dimensional Holstein model, which describes an itinerant particle that couples to an Einstein oscillator on each molecular moiety\cite{Holsteina,Holsteinb}. The use of the one-dimensional Holstein model to model charges in conjugated polymer is easily motivated. This model was originally introduced by Holstein to model charges in molecular aggregates and a polymer is simply a one-dimensional chain of covalently bonded monomers. (Indeed, the Su-Schrieffer-Heeger model\cite{Su} can be mapped onto the Holstein model\cite{Campbell}.) Similarly, a Frenkel exciton is a tightly bound electron-hole pair which hops from monomer to monomer via the exciton transfer integral, entirely analogously to an exciton on a one-dimensional J-aggregate\cite{Tozer,Yamagata,Barford2005}. Both charges and Frenkel excitons in conjugated polymers couple to the local normal mode associated with the high frequency C-C bond stretch.

The Holstein model has also been used to investigate polaron and bipolaron formation in high-temperature superconductors, and consequently a significant amount of effort has been expended investigating this model, both for ordered\cite{Capone,Jeckelmann,Wellein,Romero} and disordered\cite{bronold1,bronold2,alvermann,hague,berciu1, berciu2} systems. Here, we focus on the adiabatic and large-polaron regimes (which will be defined in Section \ref{Se:3}) in one-dimensional systems, as these are relevant to conjugated polymers. We are also interested in understanding how the fully quantized Holstein model reproduces the predictions of the classical, adiabatic (i.e., Born-Oppenheimer) limit, in which the nuclear displacements are treated classically. It is well known that in the classical limit the Holstein model predicts a `self-localized' polaron on a uniform chain. We will show that this prediction is reproduced by the quantum model in the adiabatic limit and in the limit of vanishing disorder.

This paper is organized as follows. The next section introduces the disordered Holstein model. We solve this by the Density Matrix Renormalization Group method, so a brief description of our application of this method follows.  The well-known solution of the Holstein method in the Born-Oppenheimer limit is then outlined, followed by a brief review of Anderson localization of free particles in one dimension. Section \ref{Se:3} contains our results, culminating in a proposed phase diagram of the disordered Holstein model in Section \ref{Se:4}. We conclude in Section \ref{Se:5}.


\section{Model and Methods}

\subsection{Holstein Model}

The model used in this work is the real-space quantum Holstein Hamiltonian\cite{Holsteina,Holsteinb}. On a linear chain of $N$ sites it is defined as
\begin{eqnarray}\label{Eq:1}
 \nonumber
  \hat{H} = &&-\sum_{n=1}^{N-1}t_{n}( \hat{a}_{n+1}^{\dagger}\hat{a}_{n}+\hat{a}_{n}^{\dagger}\hat{a}_{n+1})
  -g\hbar\omega\sum_{n=1}^{N}(\hat{b}_{n}^{\dagger}+\hat{b}_{n})\hat{N}_{n}
  \\
&& + \hbar\omega\sum_{n=1}^{N}\left(\hat{b}_{n}^{\dagger}\hat{b}_{n} + \frac{1}{2}\right),
\end{eqnarray}
where $\hat{a}_{n}^{\dagger}$ ($\hat{a}_{n}$) creates (destroys) a particle (e.g., a charge or Frenkel exciton) on site $n$, $\hat{N}_n= \hat{a}^{\dagger}_n\hat{a}_n$, and $\hat{b}_{n}^{\dagger}$ ($\hat{b}_{n}$) creates (destroys) an Einstein phonon of energy $\hbar\omega$ on site $n$. $g$ is the particle-phonon coupling parameter and $t_{n}$ is the particle transfer integral between sites $n$ and $(n+1)$, which is taken to be a Gaussian random variable with a mean $t_0$ and standard deviation $\sigma$.
We consider a single particle on a linear chain with open boundary conditions.


To find the low-lying eigenstates of the Holstein Hamiltonian we have used the Density Matrix Renormalization Group method\cite{White1992,White1993}. The bare particle-phonon basis on each site is truncated via single-site optimization\cite{Zhang,Barford2002}, in which a density matrix is constructed for the site basis. The reduced site basis is then augmented with the left or right system blocks by the usual real-space method. (An alternative, pseudo-site method was used by Jeckelmann and White in their study of the Holstein model\cite{Jeckelmann}.)

For the smallest phonon frequencies we typically used 25 bare phonons per site, with $\sim 15$ renormalized states per site, $\sim 40$ block states, and $\sim 200,000$ superblock states. The relative difference in energy between the ground and first electronic states (i.e., $E(j=1,v=0)$ and $E(j=1,v=0)$, as defined in Section \ref{Se:3.2})  converges to better than one part in $10^{10}$ as a function of the DMRG convergence parameters. We checked this convergence by using over $100$ block states and over $10^6$ superblock states.
At least one finite lattice sweep was performed at the target chain size, although even with disorder we never observed  improved convergence or a different energy minimum after a second sweep. 
We also note the various independent checks of consistency:
\begin{enumerate}
\item{The calculated effective masses agree with theoretical expressions in the relevant parameter regimes (see Fig.\ 4).}
\item{The particle localization lengths calculated directly from the groundstate wavefunctions (using eqn (9)) and via the effective masses (using eqn (17)) are in good agreement with each other as a function of disorder (see Fig.\ 6).}
\item{The particle density calculated via DMRG in the adiabatic limit agrees with the analytical Born-Oppenheimer expression (see Fig.\ 8). This agreement in the most challenging regime (i.e., with disorder and strongly adiabatic) provides excellent credibility to our DMRG results.}
\end{enumerate}

\subsection{The Classical Limit}\label{Se:2.2}

We first discuss the well-known solutions to eqn (\ref{Eq:1}) on a uniform one-dimensional lattice in the classical, adiabatic (i.e., Born-Oppenheimer) limit. In this limit $\omega \rightarrow 0$ and $M\omega \rightarrow \infty$ (so that the spring constant, $K = M\omega^2$, remains constant) and the phonons are treated as classical variables. Then the Born-Oppenheimer form of the Holstein model is
\begin{eqnarray}\label{Eq:2}
 \nonumber
  \hat{H}_{BO} =&& -\sum_{n=1}^{N-1}t_{n}( \hat{a}_{n+1}^{\dagger}\hat{a}_{n} + \hat{a}_{n}^{\dagger}\hat{a}_{n+1})
  -\sqrt{2}g\hbar\omega\sum_{n=1}^{N}\tilde{Q}_n \hat{N}_{n}
  \\
    && + \frac{\hbar\omega}{2}\sum_{n=1}^{N} \tilde{Q}_n^2,
\end{eqnarray}
where $\tilde{Q}_n$ is the  dimensionless nuclear displacement, defined in terms of the dimensionfull displacement, $Q_n$, via
\begin{equation}\label{}
    \tilde{Q}_n = \left(\frac{K}{\hbar \omega}\right)^{1/2} Q_n.
\end{equation}

The general  one-particle eigenstate  of eqn (\ref{Eq:2}) is
\begin{equation}\label{}
    |\Psi\rangle = \sum_n \Psi_n |n\rangle,
\end{equation}
where $\Psi_n$ is the particle wavefunction and $|n\rangle = \hat{a}^{\dagger}_n |0\rangle$ is the ket representing the particle  on site $n$.
In the continuum  limit  eqn (\ref{Eq:2}) has the exact  solution\cite{Holsteina, Rashba}
\begin{equation}\label{Eq:5}
   \Psi_{n} = \left(\frac{\lambda}{2}\right)^{1/2}\textrm{ sech }\lambda(n-n_0),
\end{equation}
where
\begin{equation}\label{Eq:6}
    \lambda = \frac{g^2 \hbar \omega}{2t_0}.
\end{equation}
Eqn (\ref{Eq:5}) describes a polaron self-localized at $n_0$, whose
 spatial extent is
\begin{equation}\label{Eq:7}
\ell_c =   2 \lambda^{-1}= \frac{4t_0}{g^2 \hbar \omega},
\end{equation}
which is assumed to satisfy $\ell_c \gg 1$, i.e., $\lambda \ll 1$ in the large polaron regime. In addition, the equilibrium nuclear displacements satisfy $\tilde{Q}_n = \sqrt{2} g |\Psi_n |^2$ and the polaron relaxation energy (defined as the difference in energy between a free and self-trapped particle) is $\Delta E_r = t_0 \lambda^2/3$. This solution implies that in one-dimension the groundstate of the Born-Oppenheimer Holstein model is a polaron for $g>0$\cite{Rashba}.

In \S\ref{Se:3} we describe how the solutions of the quantum Holstein model (eqn (\ref{Eq:1})) approach those of the Born-Oppenheimer model (eqn (\ref{Eq:2})) in the adiabatic limit. To correctly derive this limit from the quantum model, it is expedient to recast the particle-phonon coupling term of eqn (\ref{Eq:2}) as
\begin{equation}\label{}
    -\sqrt{2g^2\hbar\omega}\sum_{n=1}^{N} \sqrt{\hbar \omega} \tilde{Q}_n \hat{N}_{n}.
\end{equation}
 The classical, adiabatic limit of the quantum model  is defined as taking $\hbar\omega/t_0 \rightarrow 0$ and the number of Einstein phonons per site $\rightarrow \infty$. In taking this limit the classical displacements must remain constant, and since for fixed spring constant $\sqrt{\hbar \omega} \tilde{Q}_n \sim Q_n$, this implies that the parameter $\sqrt{g^2\hbar\omega}$ is constant.  Thus, from eqn (\ref{Eq:6}), for a fixed $t_0$ $\lambda$ is a constant. Evidently, the condition that $\lambda$ and $t_0$ remain constant means that the polaron size and  relaxation energy become  constant as the adiabatic limit is reached.

\subsection{Particle Localization}

To show that the solutions of the quantum Holstein model approach those of the classical model in the adiabatic limit, we need to show that the quantum model predicts self-localized polarons, as described by eqn (\ref{Eq:5}). On a uniform lattice these broken symmetry solutions are not permitted in the quantum model, as they are not translationally invariant. However, such solutions are permitted if the translational symmetry is broken, e.g., by the introduction of disorder. Thus, we seek solutions to the disordered quantum Holstein model in the limits that $\hbar\omega/4t_0 = \omega' \to 0$ and $\sigma \to 0$.

To quantify the polaron self-localization, we define the localization length as twice the root-mean-square spread of the particle position,
\begin{equation}\label{Eq:14}
    \ell = 2\Delta n_{rms} = 2\sqrt{\langle n^2\rangle - \langle n\rangle^2},
\end{equation}
where
\begin{equation}\label{}
   \langle n^p\rangle = \frac{ \sum_n n^p \langle \hat{N}_n\rangle }{\sum_n  \langle \hat{N}_n\rangle},
\end{equation}
$\langle  {N}_{n}  \rangle = \langle \Psi| \hat{N}_{n} |\Psi \rangle$, and $|\Psi \rangle$ is the eigenket of eqn (\ref{Eq:1}).

Since the expectation value of the nuclear displacement mirrors the particle density, i.e., $\langle b^{\dagger}_n + b_n\rangle/\sqrt{2} = \sqrt{2} g \langle N_n \rangle$, $\ell$ is also a measure of the spread of the nuclear displacements. Thus, $\ell$ may be taken as a measure of the spatial size of the polaron. (Note that in our definition the polaron size is not determined by the particle-phonon correlation length, discussed in  Section \ref{Se:3.1}.)

Since the nuclei are static in the classical, adiabatic limit, the nuclear displacements associated with the particle are self-localized. Thus, in this limit the polaron size, $\ell = \ell_c \sim 2/\lambda$, can be regarded as the self-localization length of the particle caused by its self-trapping.

Another localization length is determined by disorder. In one-dimension disorder localizes a `free' particle (where by a `free' particle we mean a particle not coupled to the phonons). According to single-parameter scaling theory\cite{Kramer}, the localization length of a free particle subject to Gaussian random disorder is
\begin{equation}\label{Eq:9}
    \ell_d^0 \sim \left( \frac{t_0}{\sigma}\right)^{\nu(E)},
\end{equation}
where the exponent ${\nu(E)}$ is a function of the energy of the particle. At the band edge ${\nu(E)} = 2/3$, while at the band center ${\nu(E)} = 4/3$.

The band width $W = 4t_0$ and since the effective mass, $m^*$, of a particle at the bottom of a parabolic band satisfies $m^* \sim W^{-1}$, we will use the effective mass of the polaron as a proxy for its band width and assume that away from the classical limit the localization length of the polaron satisfies
\begin{equation}\label{Eq:10}
    \ell_d \sim \left(\frac{m_0t_0}{m^*\sigma}\right)^{\nu(E)},
\end{equation}
where $m_0$ is the free particle mass, i.e., $ \ell_d(m^* = m_0) = \ell_d^0$. (This prediction is confirmed in Fig.\ \ref{Fi:6}.)


\section{Results}\label{Se:3}

The results presented here are in the  adiabatic \emph{regime}, defined by $\hbar\omega/4t_0 \equiv \omega'  < 1$, with the adiabatic \emph{limit} being defined as $\omega' \to 0$, subject to  $\lambda$ (defined in eqn (\ref{Eq:6})) and $t_0$ being constant. We also focus on the large polaron regime, defined by $\lambda \lesssim 1$. Unless otherwise states, all the calculations were performed on $60$ site chains with open boundary conditions. We first consider uniform chains before introducing disorder in Section \ref{Se:3.3}.

\subsection{Particle-Phonon Correlation Function}\label{Se:3.1}

\begin{figure}
\centering
\includegraphics[scale = 0.30]{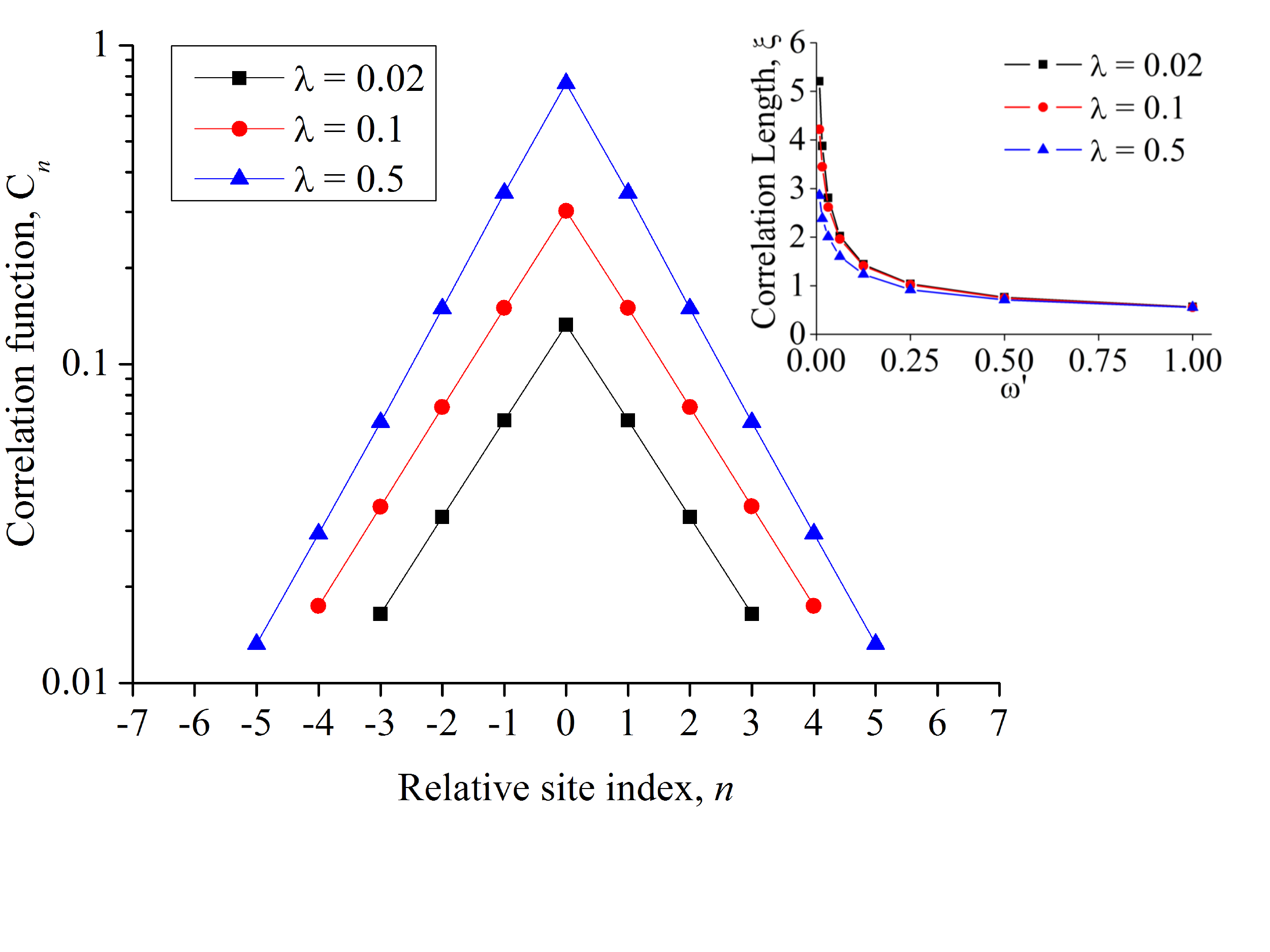}
\caption{The particle-phonon correlation function, $C_n$, defined by eqn (\ref{Eq:12}), for various $\lambda$ values with $ \omega' =  \hbar \omega/4t_0 = 0.125$. The inset shows that the correlation length, $\xi$, is virtually independent of $\lambda$, but increases as $\omega'$ decreases.}
\label{Fi:1}
\end{figure}

\begin{figure}
\centering
\includegraphics[scale = 0.30]{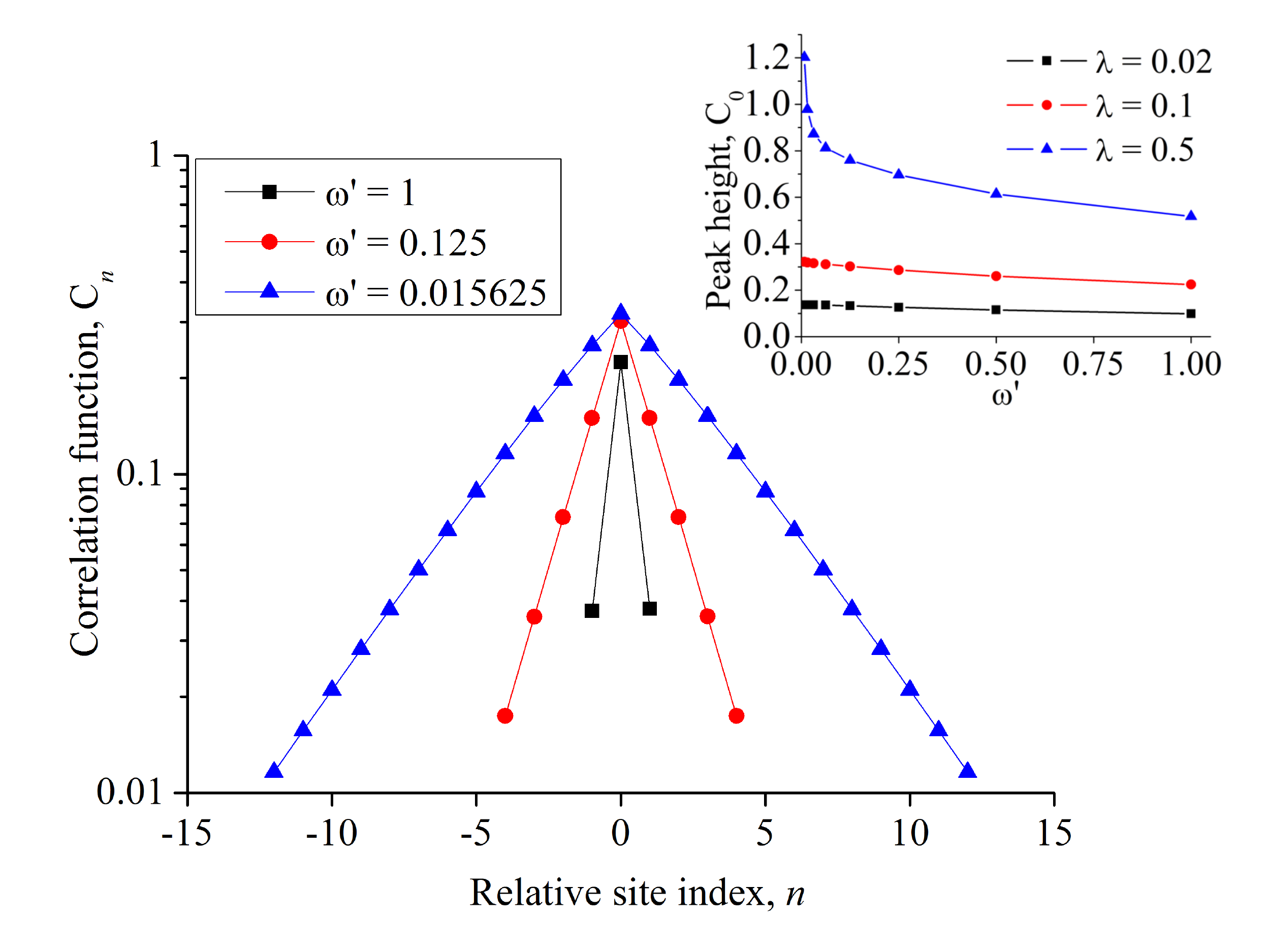}
\caption{The particle-phonon correlation function, $C_n$, defined by eqn (\ref{Eq:12}), for various $\omega'$ values with $\lambda = 0.1$. The inset shows that the peak height, $C_0$,  increases as  $\lambda$ increases.}
\label{Fi:2}
\end{figure}

The nuclear displacement associated with the polaron is most easily illustrated via the particle-phonon correlation function, $C_n$, defined by\cite{Hoffmann}
\begin{equation}\label{Eq:12}
    C_n = \frac{1}{N}\sum_m\ \frac{\langle \hat{N}_m(\hat{b}^{\dagger}_{m+n} + \hat{b}_{m+n})\rangle}{\sqrt{2}\langle \hat{N}_m\rangle}.
\end{equation}
This function correlates the nuclear displacement $n$ sites away from the instantaneous position of the particle. Fig.\ \ref{Fi:1} illustrates $C_n$ for  different $\lambda$ values for a value of $\omega' = 0.125$. The  particle-phonon correlation decays exponentially, with a correlation length, $\xi$,  shown in the inset. As expected, the correlation length increases as $\omega' \to 0$, because in this limit the phonons respond infinitesimally slowly to the particle. Fig.\ \ref{Fi:2} illustrates $C_n$ for different $\omega'$ values for a value of $\lambda = 0.1$, while the inset shows the value of $C_0$. As expected, as $\lambda$ increases (and hence the particle-phonon coupling increases) $C_0$ also increases.

These figures illustrate that the correlation function has a non-zero expectation value for all values of $\lambda$ and $\omega'$ in the adiabatic and large polaron regime, implying a self-trapped polaron. This is in accord with the theorems that state that the groundstate of the uniform Holstein model does not exhibit a phase transition\cite{Spohn, Gerlach}. To understand how the polaron becomes self-localized as a consequence of symmetry breaking, we next consider its effective mass.

\subsection{Effective Mass on a Uniform Chain}\label{Se:3.2}

We calculate the ratio of the polaron effective mass to the free particle mass, $m^*/m_0$, via,
\begin{equation}\label{Eq:11}
\frac{m^{*}}{m_0} = \frac{E_0(j=2)-E_0(j=1)}{E^*(j=2,v=0)-E^*(j=1,v=0)},
\end{equation}
where $E^*(j,v=0)$ is the energy of the lowest vibrational state of the $j$th pseudomomentum state.  We determine these states by calculating both the density, $\langle {N}_{n}\rangle$, and the transition density,  $\langle \Psi| \hat{a}_{n}^{\dagger}|0 \rangle$, where $|\Psi \rangle$ is the eigenket of eqn (\ref{Eq:1}) and $|0 \rangle$ is the particle and phonon vacuum. The absence of  nodes in both the density and transition density implies the $(j = 1, v=0)$ state, while one node in the density and no nodes in the transition density implies the $(j = 2, v=0)$ state. ($E_0$ is the free particle energy.)

\begin{figure}
\centering
\includegraphics[scale=0.30]{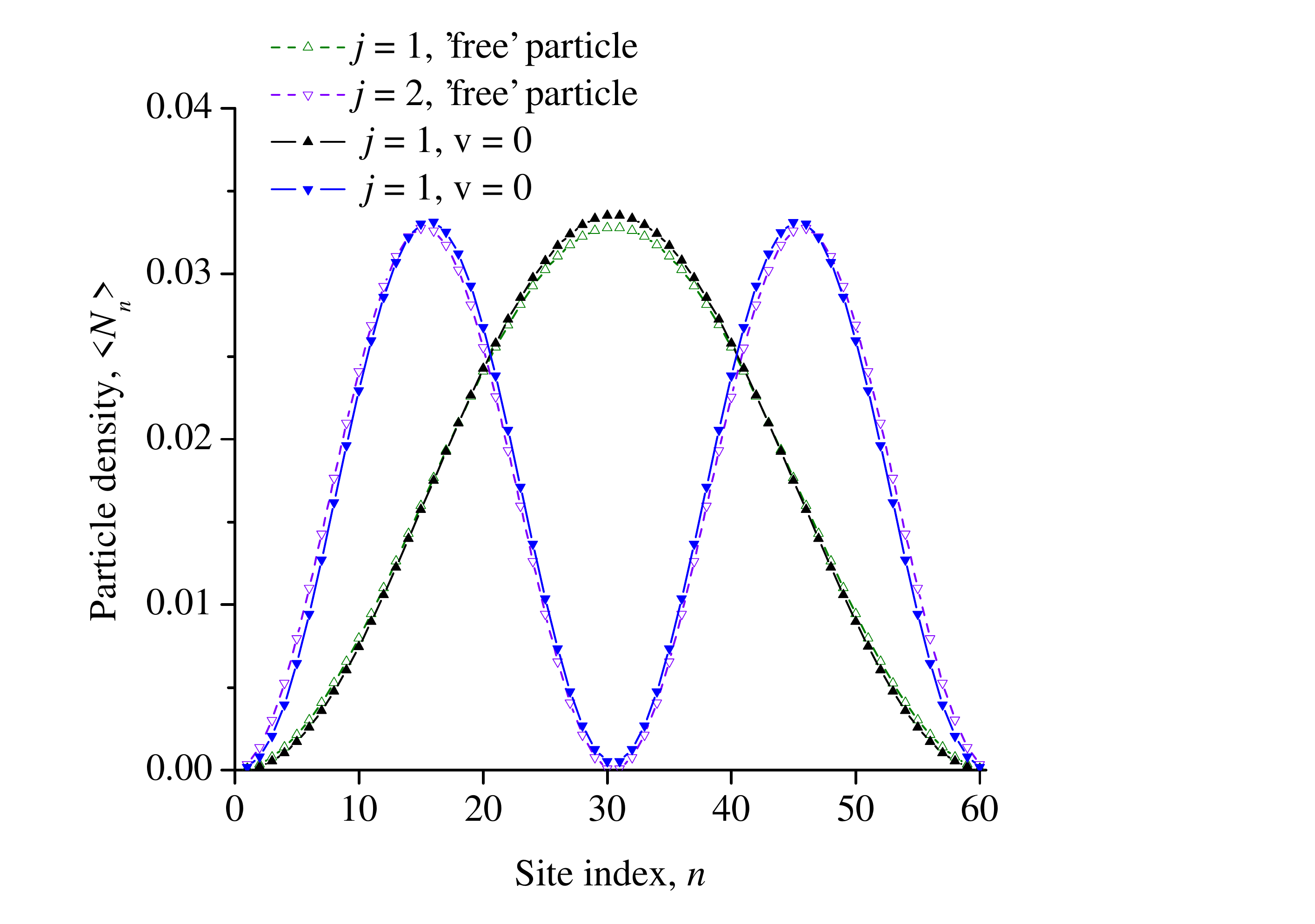}
\caption{The particle density, $\langle \Psi| \hat{N}_{n} |\Psi \rangle$, of the  $(j = 1, v=0)$ and $(j = 2, v=0)$ states. $\lambda = 0.444$ and $\omega' = 0.035$, which are appropriate model parameters for the normal mode associated with the C-C bond stretch in the conjugated polymer, poly(para-phenylene)\cite{Barford2005}. The dashed curves with open symbols are the particle densities of free particles (i.e., $\lambda = 0$) given by eqn (\ref{Eq:13b}).}
\label{Fi:3}
\end{figure}

The particle density of the  $(j = 1, v=0)$ and $(j = 2, v=0)$ states are shown in Fig.\ \ref{Fi:3}. Also shown is the  `free' particle density, given by
\begin{equation}\label{Eq:13b}
    \langle N_n \rangle = \left(\frac{2}{N+1}\right)\sin^2 \left(\frac{\pi j n}{N+1}\right).
\end{equation}
As previously observed\cite{Jeckelmann}, for non-zero values of the particle-phonon coupling, the particle is repelled by the ends of the lattice, indicating a self-trapped polaron. However, for the parameters chosen here (i.e., $\lambda = 0.444$ and $\omega' = 0.035$) this effect is rather small, because  -- as indicated in the inset of Fig.\ \ref{Fi:1} -- the particle-phonon correlation function decays rapidly for these parameters.

\begin{figure}
\centering
\includegraphics[scale=0.30]{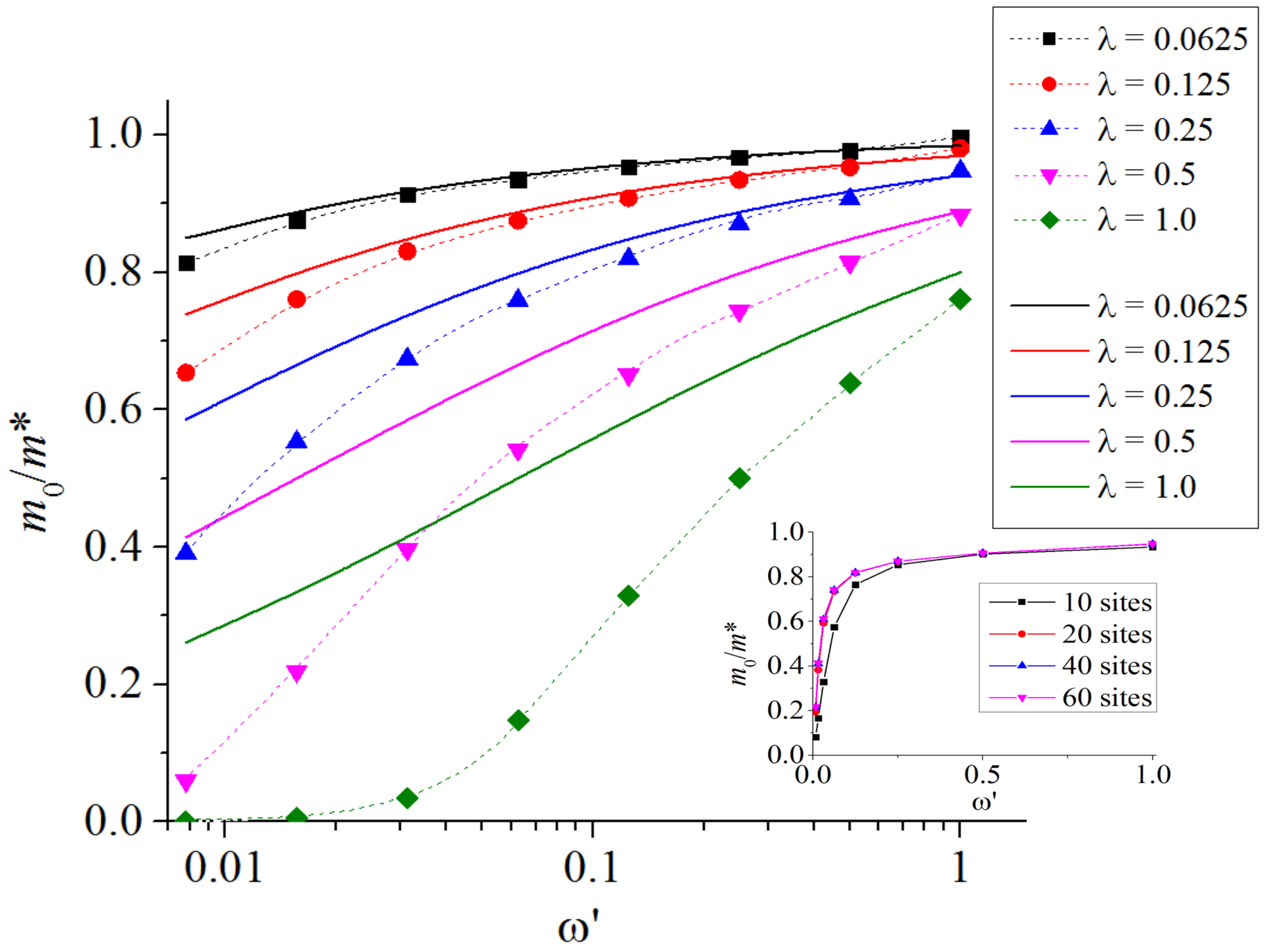}
\caption{The polaron inverse effective mass, $m_0/m^*$, versus $\omega'$ for fixed $\lambda$. The symbols and dotted curves are the DMRG calculations determined via eqn (\ref{Eq:11}). The solid curves are determined via eqn (\ref{Eq:13})\cite{Capone}, which is valid for $\omega' \ll 1$ and $\lambda \ll 1$. The inset shows $m_0/m^*$ versus $\omega'$ for $\lambda=0.25$ for chains lengths of 10, 20, 40 and 60 sites.}
\label{Fi:4}
\end{figure}

Fig.\ \ref{Fi:4} shows the inverse effective mass of the polaron versus $\omega'$ for fixed values of $\lambda$. Evidently, in the adiabatic limit the effective mass diverges. This is in agreement with the weak-coupling perturbation result\cite{Capone}
\begin{equation}\label{Eq:13}
    \frac{m^*}{m_0} = 1 + \frac{\lambda}{4\sqrt{\omega'}},
\end{equation}
which is valid for $\omega' \ll 1$ and $\lambda \ll 1$.

\begin{figure}
\centering
\includegraphics[scale=0.30]{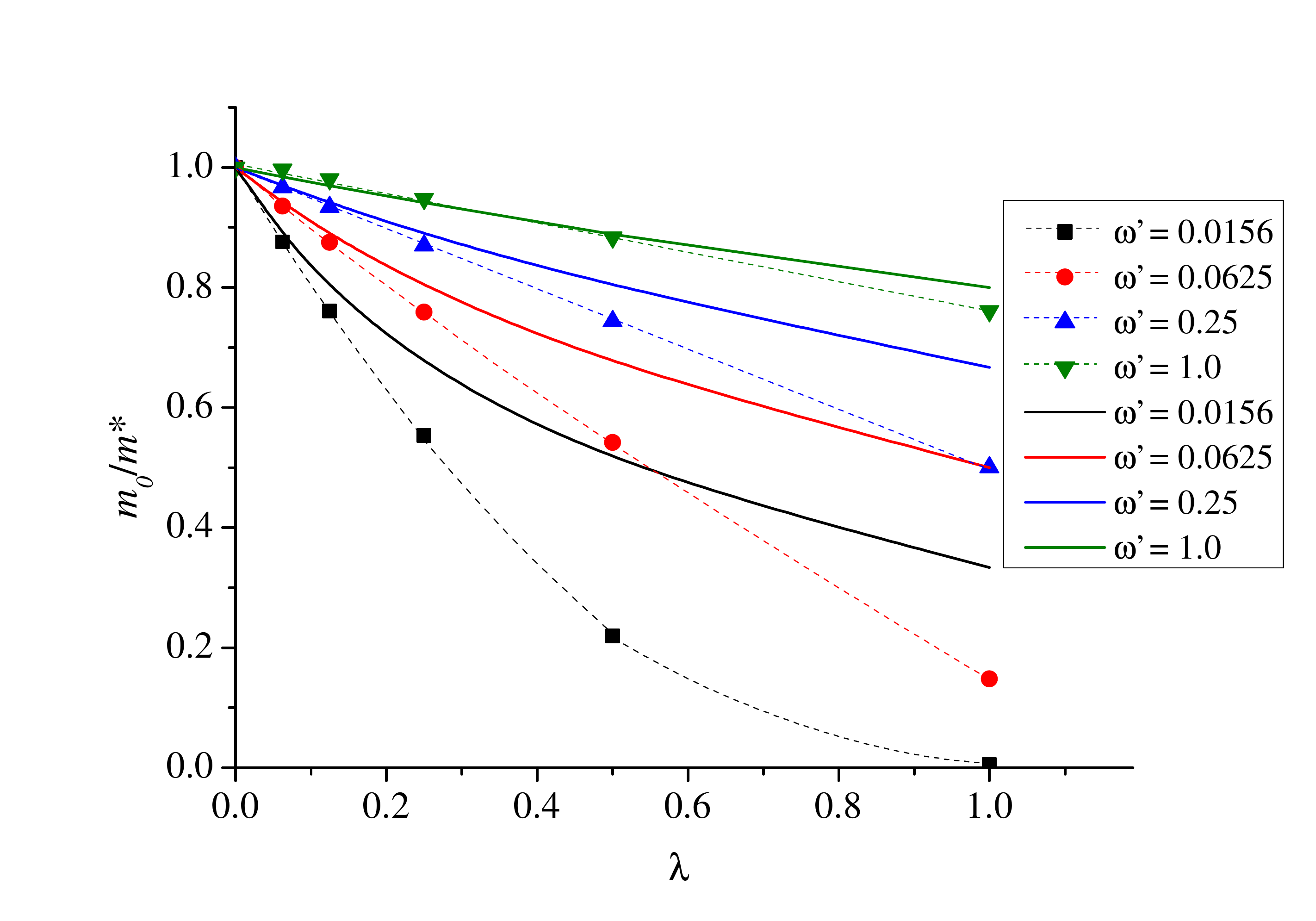}
\caption{The polaron inverse effective mass, $m_0/m^*$, versus $\lambda$ for fixed $\omega'$. The symbols and dotted curves are the DMRG calculations. The solid curves are determined via eqn (\ref{Eq:13})\cite{Capone}.}
\label{Fi:5}
\end{figure}

Similarly, Fig.\ \ref{Fi:5} shows the inverse effective mass  versus $\lambda$ for fixed values of $\omega'$. Again, the effective mass diverges as the particle-phonon coupling increases. Both figures show that for small $\lambda$ the results are in good agreement with  weak-coupling perturbation theory, eqn (\ref{Eq:13}), for a wide range of $\omega'$.

We checked for finite size effects by calculating $m_0/m^*$ versus $\omega'$ for a fixed $\lambda$ for chains lengths of $10-60$ sites. As the inset of Fig.\ \ref{Fi:4} shows, finite size effects become negligible for chains lengths over $20$ sites.

\subsection{Localization on a Disordered Lattice}\label{Se:3.3}

\begin{figure}
\centering
\includegraphics[scale=0.35]{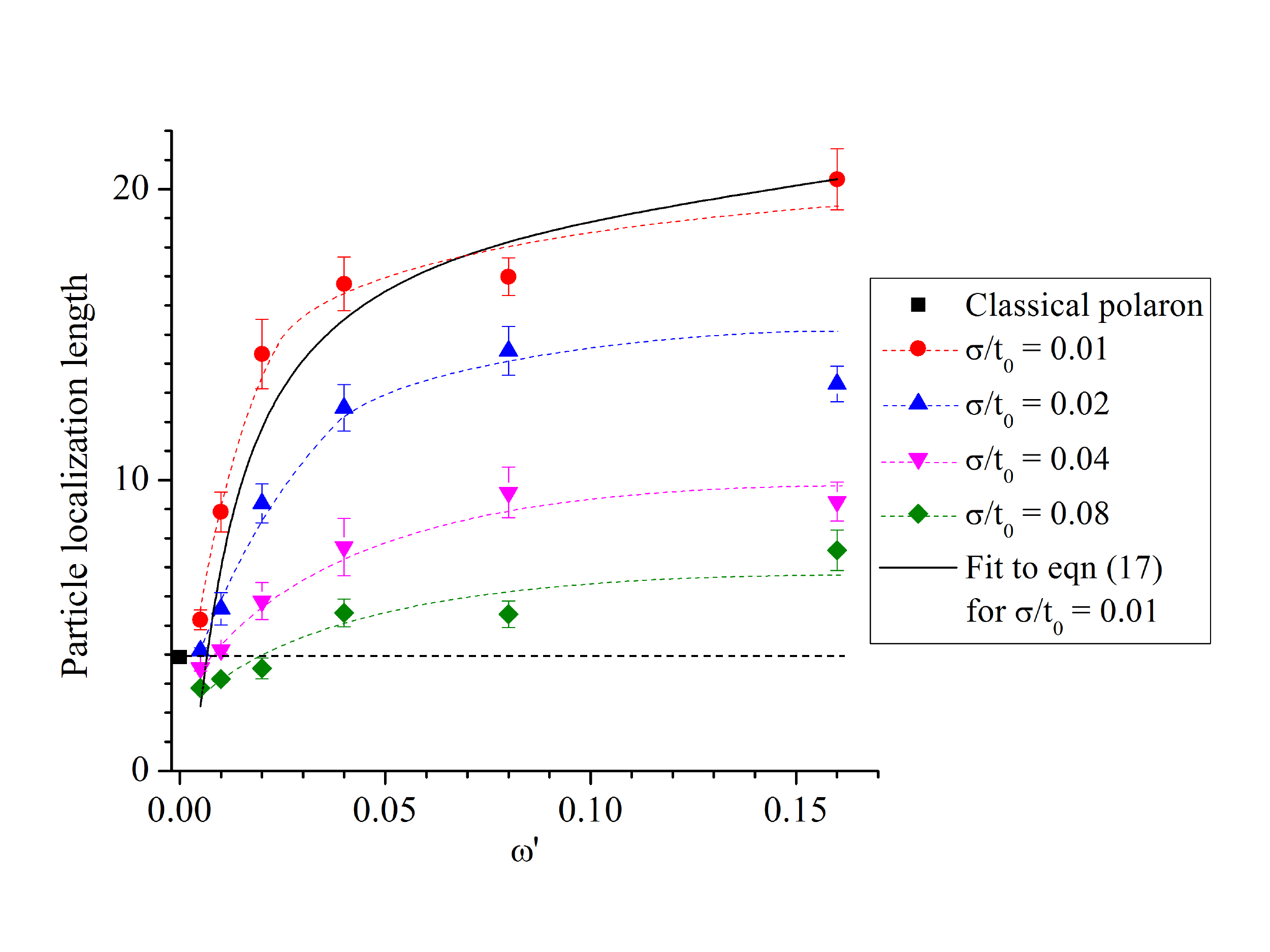}
\caption{The particle localization length, defined by eqn (\ref{Eq:14}), versus $\omega'$ for different $\sigma/t_0$. $\lambda = 0.444$. For weak disorder this is an interpolation between eqn (\ref{Eq:10}) in the non-adiabatic regime and eqn (\ref{Eq:7}) in the adiabatic limit, as indicated by the fit to eqn (\ref{Eq:16}) for $\sigma/t_0 = 0.01$ (solid curve) and the classical polaron size (square and horizontal dashed line). (Note, $3.91$ is twice the root-mean-square classical polaron size on a lattice. In contrast, the continuum size, $2\lambda^{-1} = 4.50$ when $\lambda = 0.444$.)}
\label{Fi:6}
\end{figure}

We now consider the role of disorder on polaron localization. Fig.\ \ref{Fi:6} shows the polaron localization length, $\ell$, determined via eqn (\ref{Eq:14}), as a function of $\omega'$ for different degrees of disorder. The value of $\lambda$ is $0.444$, which is appropriate for Frenkel excitons coupled to the normal mode associated with the C-C bond stretch in the conjugated polymer, poly(para-phenylene)\cite{Barford2005}. Each data point is an ensemble average over 20 realizations of the disorder.

If the disorder is weak enough so that the disorder-induced localization length of the free particle $\ell_d^0$, given by eqn (\ref{Eq:9}), is larger than the classical polaron size (i.e., the self-trapped localization length $\ell_c$, given by eqn (\ref{Eq:7})), then the localization length decreases as $\omega'$ decreases, because  the effective mass increases. Using eqn (\ref{Eq:10}) and the calculated effective masses, this prediction is confirmed in Fig.\ \ref{Fi:6}, which shows a fit to
\begin{equation}\label{Eq:16}
    \ell(\omega') = \ell(\omega' = 0.16)\left(\frac{m^*(\omega')}{m^*(\omega'=0.16)}\right)^{2/3}
\end{equation}
for $\sigma = 0.01$.

Clearly, $\ell(\omega')$ defined by eqn (\ref{Eq:16}) vanishes as $\omega'$ vanishes. Thus, when the particle localization length, $\ell$, is of the order of the classical polaron size, $\ell_c \sim 2/\lambda$, there is a cross-over from disorder-induced particle localization to self-localization induced by self-trapping. The localization length, $\ell$, is therefore an interpolation between eqn (\ref{Eq:10}) with
$m^* = m_0$ for $\omega' \gtrsim 1$ and eqn (\ref{Eq:7}) as $\omega' \to 0$. This is confirmed by our results shown in Fig.\ \ref{Fi:6}, which shows that the localization lengths for $\sigma/t_0 = 0.01$, $0.02$, and $0.04$ converging on the classical polaron size in the  adiabatic limit.

In summary, as the adiabatic limit is approached from above there is a cross-over  from a regime where the polaron size is determined by the disorder-induced localization of the particle to a regime where the polaron is self-localized and its size is determined solely by $\lambda$. This cross-over is a function of $\sigma$ and becomes a step function at $\omega'=0$ as $\sigma \to 0$.

Finally, for disorder so large that the free particle disorder-induced localization length is comparable to or smaller than the classical polaron size (i.e., $(t_0/\sigma)^{2/3} \lesssim 2/\lambda$), then disorder determines the polaron size, even in the adiabatic limit. This is indicated in Fig.\ \ref{Fi:6} for $\sigma/t_0 = 0.08$.

\section{Phase Diagram}\label{Se:4}

\begin{figure}
\centering
\includegraphics[scale=0.30]{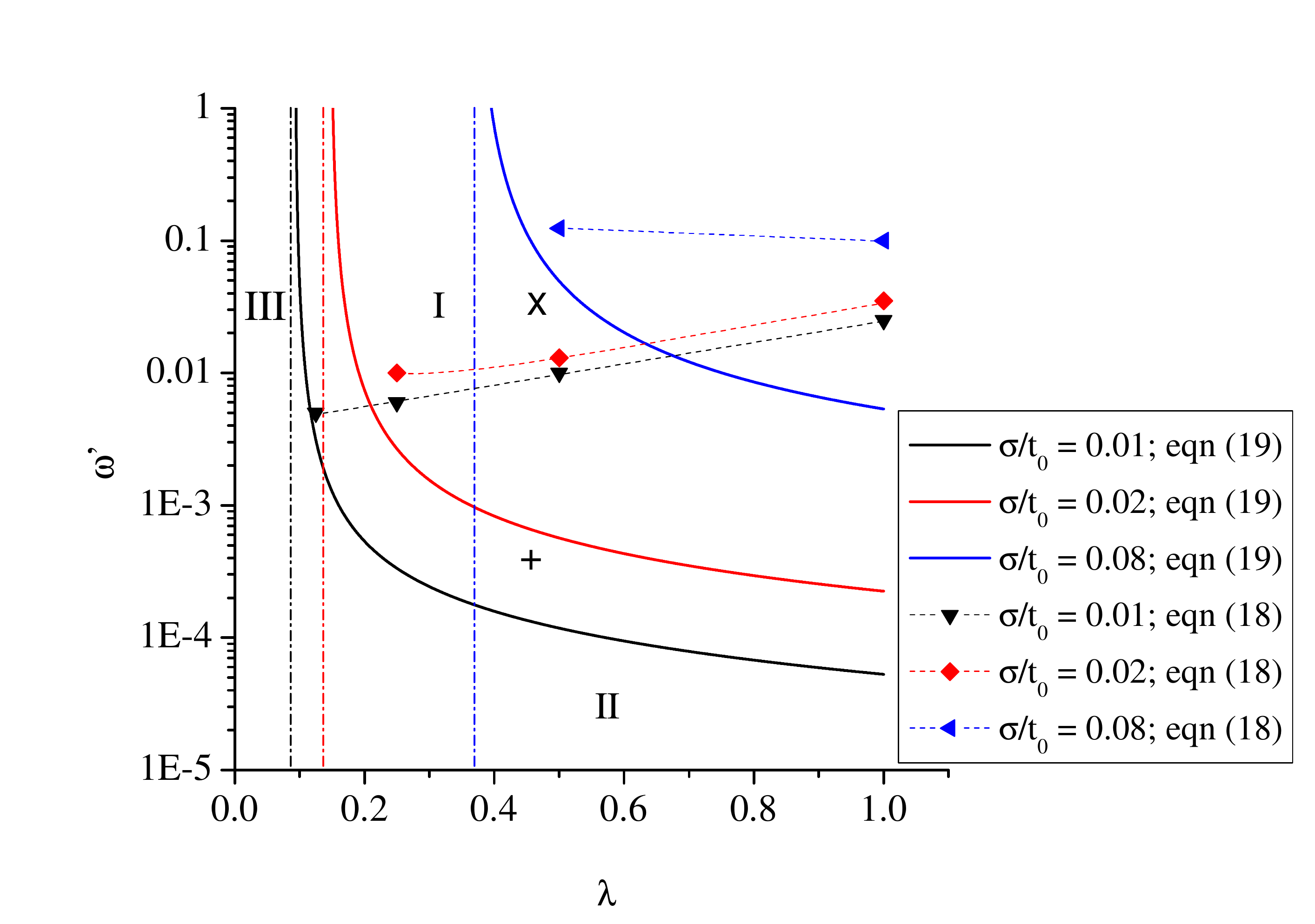}
\caption{Qualitative phase diagram of the disordered Holstein model for different disorders, $\sigma/t_0$.
Regime I: To the right of the vertical dash-dot lines and above the solid and dotted curves is disorder-induced localization.
Regime II: To the right of the vertical dash-dot lines and below the solid and dotted curves is polaron self-localization.
Regime III: to the left of the vertical dash-dot lines is strong disorder.
The boundary of regimes I and II is determined numerically via eqn (\ref{Eq:18a}) and Fig.\ \ref{Fi:4} (symbols and dotted curves), and by eqn (\ref{Eq:18}) (solid curves). The cross-over to regime III from regimes I and II is determined by $\ell_d^0 < \ell_c$.
These regimes are indicated for $\sigma/t_0 = 0.01$. The $\times$ and $+$ symbols indicate the parameters used for the calculation of the particle density shown in Fig.\ \ref{Fi:8}.}
\label{Fi:7}
\end{figure}

We can use the insights of the last section to construct a qualitative phase diagram for the disordered Holstein model in the adiabatic and large polaron regimes. We identify three regimes:
\begin{itemize}
\item{Regime I, defined by $\ell_c < \ell_d \le \ell_d^0$ (where $\ell_c$, $\ell_d$, and $\ell_d^0$ are given by eqns (7), (12), and (11), respectively). In this regime disorder and electron-phonon coupling are weak enough that the particle is localized by disorder, albeit enhanced  by its increased effective mass. Here $\ell = \ell_d$. The failure of the classic limit to predict the polaron size indicates a break-down of the Born-Oppenheimer approximation in this regime.}
\item{Regime II, defined by $\ell_d < \ell_c < \ell_d^0$. This is the adiabatic limit of regime I, such that $m^* \to \infty$ and $\ell_d \to 0$. Here the polaron is self-localized and its size is determined by the localization of the particle self-trapped by its own nuclear displacements. Thus, $\ell = \ell_c$.}
\item{Regime III, defined by $\ell_d \le \ell_d^0 < \ell_c$. This is the regime of strong disorder, and thus $\ell = \ell_d$.}
\end{itemize}
Regime I (where the localization length is larger than the internal size of the polaron) and regime III (where the localization length is smaller than the internal size of the polaron) are discussed in ref\cite{bronold2} at the mobility edge of three-dimensional systems.

The cross-over from regimes I and II occurs at $\ell_c = \ell_d$. Using eqn (\ref{Eq:7}) and eqn (\ref{Eq:10})  the cross-over occurs at
\begin{equation}\label{Eq:18a}
  \frac{m^*}{m_0} = \left(\frac{t_0}{\sigma}\right)\left(\frac{\lambda}{2}\right)^{3/2}.
\end{equation}
Using the calculated values of $m^*/m_0$ shown in Fig.\ \ref{Fi:4} and the cross-over to regime III from regimes I and II  determined via $\ell_d^0 < \ell_c$, the phase diagram is shown in Fig.\ \ref{Fi:7}.
Furthermore, the perturbative expression for $m^*/m_0$ given by eqn (\ref{Eq:13}) can be used in eqn (\ref{Eq:18a}), giving
\begin{equation}\label{Eq:18}
    \left(1 + \frac{\lambda}{4\sqrt{\omega'}}\right) = \left(\frac{t_0}{\sigma}\right)\left(\frac{\lambda}{2}\right)^{3/2}.
\end{equation}
The cross-over determined by eqn (\ref{Eq:18}) is also shown in Fig.\ \ref{Fi:7}.
Compared to the numerical results, obtained from eqn (\ref{Eq:18a}) and Fig.\ \ref{Fi:4}, the analytical result, obtained from (\ref{Eq:18}), underestimates the domain of the Born-Oppenheimer solutions (i.e., regime II), because eqn (\ref{Eq:13}) underestimates $m^*$.

\begin{figure}
\centering
\includegraphics[scale=0.30]{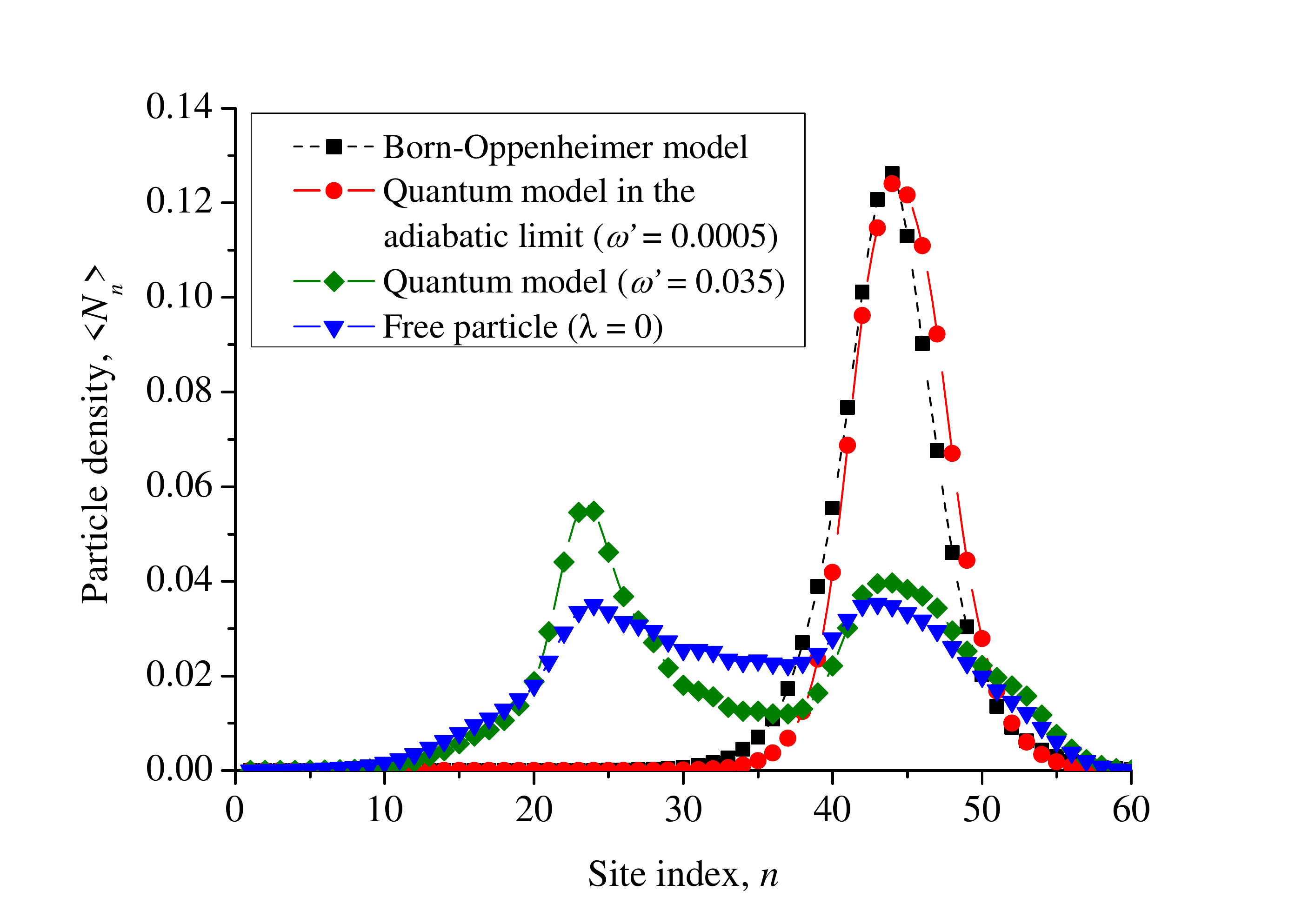}
\caption{The groundstate particle densities determined from the quantum disordered Holstein model (solid curves; $\omega' = 0.035$ (diamonds), $\omega' = 0.0005$ (circles)) and the Born-Oppenheimer disordered Holstein model (squares and dashed curve). This shows that in adiabatic limit the quantum model reproduces the solution of the Born-Oppenheimer model, given by eqn (\ref{Eq:5}). $\lambda = 0.444$ and $\sigma/t_0 = 0.02$.
The relevant parameters are shown as $\times$ and $+$ symbols in Fig.\ \ref{Fi:7}.
Also shown is the free particle density (i.e., $\lambda = 0$) for the same disorder (triangles). The dimensionless nuclear displacements satisfy $\langle \tilde{Q}_n \rangle = \sqrt{2} g \langle N_n \rangle$.
The values of $\lambda = 0.444$ and $\omega' = 0.035$ are appropriate model parameters for the C-C bond stretch in the conjugated polymer, poly(para-phenylene), showing that the Born-Oppenheimer approximation fails for this system.}
\label{Fi:8}
\end{figure}

The cross-over from a disorder-induced localized polaron wavefunction (regime I) to a self-localized polaron wavefunction (regime II) is indicated by Fig.\ \ref{Fi:8}, which shows the particle density calculated from the quantum Holstein model in these two regimes. Also shown is the classical polaron wavefunction calculated from the Born-Oppenheimer Holstein model. The particle density in the adiabatic limit of the quantum Holstein model is in good agreement with the corresponding classical density and to that of eqn (\ref{Eq:5}). As the adiabatic limit is approached the particle density smoothly interpolates to that of the classical density. Fig.\ \ref{Fi:8} also shows the particle density of a free particle, which closely resembles the particle density for the relevant Frenkel exciton poly(para-phenylene) parameters, indicating that this system is not in the Born-Oppenheimer limit.


\section{Conclusions}\label{Se:5}

We have solved the disordered Holstein model via the DMRG method to investigate the combined roles of electron-phonon coupling and disorder on the localization of a single charge or exciton. The parameter regimes chosen, namely the adiabatic regime, $\hbar\omega/4t_0 = \omega'  < 1$ and the `large' polaron regime, $\lambda < 1$, are applicable to most conjugated polymers. We showed that as a consequence of the polaron effective mass diverging in the adiabatic limit (defined as  $\omega' \to 0$ subject to fixed $\lambda$) self-localized, symmetry breaking solutions are predicted by the quantum Holstein model for infinitesimal disorder -- in complete agreement with the predictions of the Born-Oppenheimer Holstein model. (This is regime II of the phase diagram shown in Fig.\ \ref{Fi:7}.)

For many parts of the ($\omega'$, $\lambda$) parameter space, however, self-localized Born-Oppenheimer solutions are not expected. If $\omega'$ is not small enough and $\lambda$ is not large enough, then  the polaron is predominately localized by Anderson disorder, albeit more than for a free particle, because of the enhanced effective mass. (This is regime I of the phase diagram shown in Fig.\ \ref{Fi:7}.) Alternatively, for very small particle-nuclear coupling ($\lambda \ll 1$) the disorder-induced localization length is always smaller than the classical polaron size, $2/\lambda$, so that disorder always dominates. (This is regime III of the phase diagram shown in Fig.\ \ref{Fi:7}.)

It is expected that the relevant parameter regimes for the normal mode associated with the high frequency C-C bond stretch in conjugated polymers are I or III. For example, for Frenkel excitons in poly(para-phenylene), $\omega' \sim 0.035$ and $\lambda \sim 0.444$, placing them in regime I for most sensible ranges of disorder. For charges, on the other hand, $\omega' \sim 0.062$ and $\lambda \sim 0.198$ placing them in regime III for most sensible ranges of disorder. Thus, the self-localized, Born-Oppenheimer solutions of the Holstein model are not relevant for this normal mode. However, torsional modes are typically 20 times smaller in frequency\cite{tretiak}, and so for these modes Born-Oppenheimer solutions may be appropriate, provided that their coupling to the electronic degrees of freedom is strong enough.



\begin{thebibliography}{98}
\bibitem{footnote1} The term `self-trapping' is ambiguous. Here we follow the customary literature definition where it means a particle associated with nuclear displacements, which together are delocalized in space. In this paper we distinguish `self-trapping' from `self-localization' (or `auto-localization'), where `self-localization' refers to a polaron in the classical, adiabatic limit which is localized in space (see Section \ref{Se:2.2}).
\end{thebibliography}
\end{document}